\documentclass[a4paper,11pt]{article}
\usepackage[dvipsnames]{xcolor}
\usepackage{graphicx,hyperref,natbib,tikz}
\definecolor{DarkerRed}{HTML}{A03030}
\definecolor{DarkerBlue}{HTML}{004080}
\definecolor{DarkerGreen}{HTML}{008040}
\hypersetup{colorlinks=true,linkcolor=DarkerRed,urlcolor=DarkerBlue,citecolor=DarkerGreen}

\begin{document}
\title{Milky Way evolution on a human timescale}
\author{\textsc{Eugene} and \textsc{Neige}}
\date{1 April 2026}
\maketitle

\begin{abstract}
How do galaxies form and evolve? This is one of the most puzzling questions in astronomy. 
Galaxy assembly takes place throughout the entire history of the Universe, but our understanding of it is hampered by the unfortunate fact that we can only observe galaxies at a single moment in time.
Here, we use archival data of decades-long monitoring of the Milky Way to examine some of its key characteristics, namely the mass of its central black hole, the pattern speed of the bar, and the distance from the Sun to the Galactic centre.
We find a surprisingly fast evolution of these three properties on a timescale of only a few decades, and speculate that it might be driven by shared physical processes.
\end{abstract}

\section{Introduction}

Astronomy has always been associated with the vast spatial and temporal scales of the Universe. The dynamical range of astrophysical phenomena is so large that it is often sufficient to get an order-of-magnitude estimate of an effect without detailed modelling\footnote{Some theorists even go as far as declaring that ``it's ok to make an error in the order of magnitude, so long as the order of the order of magnitude is correct'' (V.Lipunov, priv.comm.)}. Nevertheless, observational data become more precise with time, demanding an equally rigorous theoretical analysis and sometimes leading to dramatic shifts in our understanding of the Universe, e.g., arising from the JWST observations of high-redshift galaxies \citep{Adamo2025}. In this work, we bring this ongoing revolution closer to home (i.e., to our own Galaxy), by carefully reassessing recent measurements of key properties of the Milky Way and demonstrating that these properties do evolve surprisingly rapidly.

Although most interesting processes (at least for the intended audience of this research note) indeed occur on astronomically large timescales, this is by no means a universal law. Examples of cosmic phenomena measurable on a timescale of a few years, and recognized by appropriate N-awards, include the deceleration of binary pulsars due to gravitational waves \citep{Hulse1975, Taylor1989}, the discovery of exoplanets thanks to the time-varying radial velocity signal \citep{MayorQueloz1995}, and the motion of S-stars around the supermassive black hole (SMBH) in the Galactic centre \citep{Ghez2008, Genzel2010}. And indeed, we will start with the last example.

\section{The mass of the Milky Way's central black hole}  \label{sec:mbh}

\begin{figure}[p]
\refstepcounter{figure}\label{fig:bh_mass}
\includegraphics{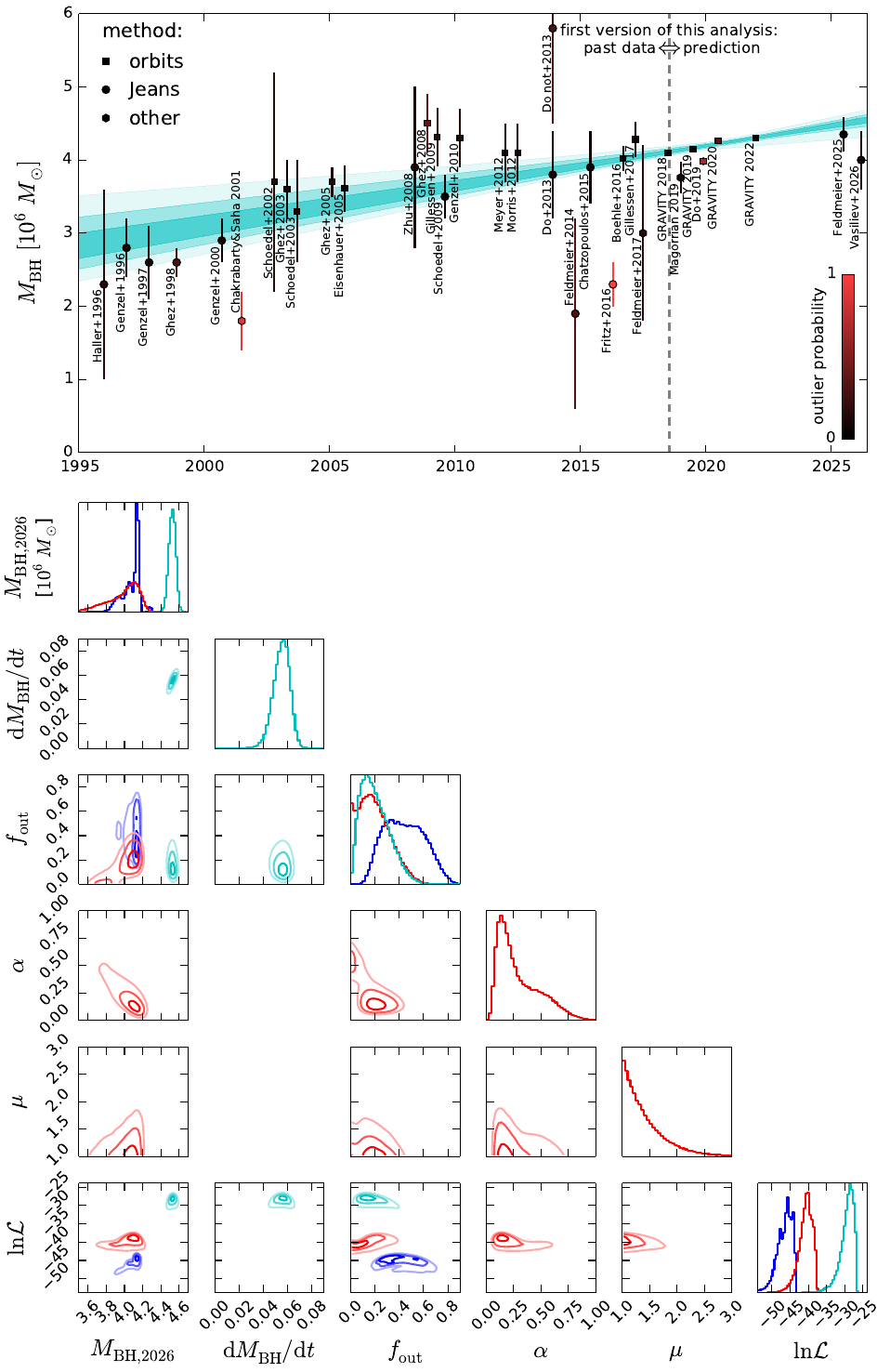}
\raisebox{105mm}[0mm][0mm]{\makebox[0mm][l]{\parbox{16cm}{\parshape=24
39mm 120mm
39mm 120mm
39mm 120mm
39mm 120mm
64mm 95mm
64mm 95mm
64mm 95mm
64mm 95mm
64mm 95mm
89mm 70mm
89mm 70mm
89mm 70mm
89mm 70mm
89mm 70mm
114mm 45mm
114mm 45mm
114mm 45mm
114mm 45mm
114mm 45mm
139mm 20mm
139mm 20mm
139mm 20mm
139mm 20mm
139mm 20mm
\textbf{Figure \arabic{figure}.} Upper panel: evolution of Sgr A$^\star$ mass in the last 30 years.\\
Cyan shaded region show the trend obtained by fitting the Model 3 (linear relation with outliers), as described in the text, and points are coloured by the probability of being outliers.\\[0.5mm]
Lower panels: corner plot of fit parameters in three series of models. \textcolor{blue}{Model 1 (blue)} is a constant $M_{\rm BH}$ with a significant fraction of measurements ($f_{\rm out}$) classified as outliers. \textcolor{red}{Model 2 (red)} reduces the fraction of outliers at the expense of considerably inflating the measurement errors by $\alpha$dditive and $\mu$ultiplicative factors. \textcolor{cyan}{Model 3 (cyan)} instead allows for a linear trend in $M_{\rm}$ with a slope ${\rm d}M_{\rm BH}/\rm{d}t$ and current value $M_{\rm BH, 2026}$, and has a much higher overall likelihood.\\[0.5mm]
D. Kanheman argues that the human brain is often biased to prefer narratives that are more plausible (despite being less likely). \\[0.5mm]
We scientists \mbox{are not} \mbox{immune} to those.}}}%
\raisebox{78mm}[0mm][0mm]{\makebox[0mm][l]{\hspace{45mm}\href{https://arxiv.org/abs/2503.24254}{[AD]}}}%
\raisebox{53mm}[0mm][0mm]{\makebox[0mm][l]{\hspace{45mm}\href{https://github.com/dmrowan/promoplot}{[AD]}}}%
\end{figure}

It has been suspected since 1990s that our Galactic centre harbours an SMBH. Once the suspicion reached a critical level, the N-committee recognized the gravity of the situation\footnote{pun intended.} and issued a cautionary statement about ``the discovery of a supermassive compact object'' (without explicitly calling it a black hole). Regardless of the physical nature of this object, it is instructive to look at the progress of its mass measurement over the last 30 years.

Figure~\ref{fig:bh_mass}, top panel, shows the published values of Sgr A$^\star$\footnote{a classic example of wretched naming traditions in astronomy; in every public talk, one has to explain that this object is anything but a \textit{star}.} obtained in $\sim$30 studies. 
It should not escape attention of a careful reader that there is an upward trend in these measurements (ignore for the moment the cyan shaded region).
In fact, the first version of this plot, with data up to 2019 (indicated by a dashed line), was \href{https://eugvas.net/docs/talk_nov18_mwnsc.pdf}{presented} by one of us at a conference a few years ago, and the trend was already fairly obvious back then. What is striking is that a series of much more precise measurements from the \citet{Gravity2018,Gravity2019,Gravity2020,Gravity2022} fall \textit{exactly} on the same line, making it a truly spectacular case of a successful prediction\footnote{See \citet{McGaugh2020} for a historical perspective on a range of phenomena successfully predicted by one cosmological model but not the other.\label{footnote:dm}}.

But how sure are we that this trend is real and not just a statistical curiosity caused by a few outlier points? A simple calculation shows that a constant $M_\mathrm{BH}$ is extremely unlikely: the value of $\chi^2$/d.o.f.\ exceeds 10. By contrast, a linear fit to $M_\mathrm{BH}$ gives a much lower $\chi^2$/d.o.f.\ of 3.5, which is still not great, but more plausible. However, a cursory glance at the dataset reveals that there are a few clear outliers. While a standard\footnote{and deplorable!} practice in astronomy is to prune them with some sort of $\sigma$-clipping, we can do much better by following an excellent pedagogical tutorial by \citet{Hogg2010}. Namely, we write a \textit{likelihood function} of a mixture model that accounts for the presence of outliers, possible underestimation of measurement errors, and either a constant or a linearly-varying $M_\mathrm{BH}$:
\begin{equation}
\ln \mathcal L = \sum_{i=1}^{N_\mathrm{data}} \ln \Big[ (1-\eta)\,\mathcal P_\mathrm{mod}(M_{\mathrm{BH},i}, \epsilon_{M_\mathrm{BH},i}, t) + \eta\,\mathcal P_\mathrm{out}(M_{\mathrm{BH},i}) \Big], 
\end{equation}
Here $\eta$ is a free parameter describing the probability of any given measurement $M_{\mathrm{BH},i}$ to be an outlier described by a flat distribution $P_\mathrm{out}(M_\mathrm{BH}) = 1 / (M_\mathrm{max}-M_\mathrm{min})$, and the probability of measuring a value $M_{\mathrm{BH},i}$ with an uncertainty $\epsilon_{M_\mathrm{BH},i}$ at time $t$ in the linear model for the SMBH mass evolution is
\begin{equation}
\mathcal P_\mathrm{mod}(M_{\mathrm{BH},i}, \epsilon_{M_\mathrm{BH},i}, t) =
\frac{1}{\sqrt{2\pi \big[ (\mu\,\epsilon_{M_\mathrm{BH},i})^2 + \alpha^2 \big]}}
\exp\left[-\frac{\big\{M_{\mathrm{BH},i} - [M_\mathrm{BH,2026} + A\,(t-2026)]\big\}^2}{2\, \big[ (\mu\,\epsilon_{M_\mathrm{BH},i})^2 + \alpha^2 \big]} \right] .
\end{equation}
By making $A \equiv \mathrm{d}M_\mathrm{BH}/\mathrm{d}t$ a free parameter or fixing it to zero, we can compare the likelihoods of models with a growing or a constant SMBH mass.
We may also account for a possible underestimation of measurement errors by introducing additive ($\alpha$) and multiplicative ($\mu$) factors. This likelihood function is provided to the Markov Chain Monte Carlo fitting code \texttt{emcee} \citep{ForemanMackey2013}, and the posterior distributions of parameters\footnote{In all cases, the probability of each measurement to be an outlier is given by $\mathcal P_\mathrm{out} / [\mathcal P_\mathrm{out} + \mathcal P_\mathrm{mod}(M_{\mathrm{BH},i})]$, and $f_\mathrm{out}$ is the mean value of these probabilities across the entire dataset (strongly related but not identical to $\eta$).} for different choices of models is shown in the lower part of Figure~\ref{fig:bh_mass}.

In the baseline model with a constant $M_\mathrm{BH}$ and no tweaking of measurement uncertainties, between 30 and 70\% of all measurements are declared to be outliers (blue curves), which is clearly unacceptable -- we cannot admit that half of our colleagues (including some N-laureates!) were wrong. We can amend the situation a little bit by adjusting the quoted uncertainties by $\alpha$dditive and $\mu$ultiplicative factors (red curves); however, in this case the required adjustment factos are quite large (e.g., $\alpha \simeq 0.5\times 10^6\,M_\odot$ if we wish to reduce the fraction of outliers), negating the improvements in observational technology in the past few decades (in particular, the huge increase in precision brought by the GRAVITY interferometer). On the other hand, if we abandon the null hypothesis of $A=0$ and consider a linearly growing $M_\mathrm{BH}$ (cyan curves), the log-likelihood of the model increases by more than 10 even without tweaking the errors, and only a couple of measurements are considered oddballs. We conclude that with a robust statistical analysis, there is an overwhelming evidence that the SMBH mass growth with time at a rate of $\sim 0.06\times 10^6\,M_\odot$/yr.

The idea that the SMBH mass may grow substantially in the timespan of only a few decades may seem absurd, but it remains a duty of a self-respecting scientist to at least contemplate the possibility of correctness of the above analysis, and consider its implications\footnote{We stop short of blaming the authors of previous papers for failing to do so. Perhaps they had their reasons, e.g. wanted the papers to be published, or some other equally lame excuse...}. First, we need to identify possible physical mechanisms of such a rapid mass increase. It is immediately obvious that it cannot be driven by gas accretion, as the required accretion rate exceeds the Eddington rate by many orders of magnitude, greatly surpassing even the wildest fantasies of theorists striving to explain the unusually heavy SMBH in the early Universe discovered by JWST\footnote{\citet{Marziani2025} discuss super-Eddington accretion, we understand none of it, but thought it might be relevant.} \citep[e.g.,][]{Natarajan2024}. Dark matter (DM\footnote{see footnote \ref{footnote:dm}\label{footnote:dm2}}) can be accreted without producing any electromagnetic radiation, and thus is not subject to the limitations of Eddington luminosity. However, calculations performed in the standard cold DM scenario show that the accreted mass over the Hubble time is typically a small fraction of $M_\mathrm{BH}$ \citep[e.g.,][]{Ilyin2004,Merritt2004,Peirani2008,Vasiliev2008}. But of course, cold DM is not the only possibility considered by theorists, and in the self-interacting DM scenario, haloes undergo core collapse, which could even result in a formation of a SMBH \citep{Balberg2002}. We are far from suggesting that Sgr A$^\star$ has just formed in such a process a few decades ago, as there is strong evidence of it existing and accreting gas at a high rate at least a few centuries to $10^7$ years ago, producing light echoes in the surrounding gas clouds \citep{Revnivtsev2004,Marin2023} and Fermi bubbles in the Galactic halo \citep{Su2010}. However, if the core collapse of a SIDM halo occurs just now\footnote{or rather, 28000 years ago, accounting for the light travel time.}, the dramatic increase in the central density of DM might boost the accretion rate by orders of magnitude. Whether this is sufficient to account for the observed mass growth rate is left for a follow-up study.

Having found a satisfactory physical explanation for the measured mass increase, we now consider its implications over longer timescales. For simplicity, we assume that the current growth rate stays constant, extrapolating the mass linearly, as we are clearly entitled to do\footnote{see \url{https://xkcd.com/605}. To be sure, we repeated the fits allowing for a quadratic term in $M_\mathrm{BH}(t)$, but it was consistent with zero, so we opted for a safer and most realistic assumption of linear growth.}. First, let us work out the relevant numbers: at a current growth rate, the SMBH would reach $10^7\,M_\odot$ in less than a hundred years, and $10^{11}\,M_\odot$ -- in $(1.7\pm 0.2)\times 10^6$\,yr. For context, this exceeds \href{https://en.wikipedia.org/wiki/List_of_most_massive_black_holes}{all SMBH mass estimates} that researchers dared to publish so far, and is approximately equal to the mass of the Milky Way interior the Solar orbital radius (see Figure~16 in \citealt{Hunt2025}), of which roughly half is believed to be contributed by DM.

\textit{\href{https://knowyourmeme.com/memes/i-accidentally}{I accidentally the whole Galaxy. Is it dangerous?}}\\
At first, one might think that we would still be far from disaster: the Schwarzschild radius of a $10^{11}\,M_\odot$ SMBH is only 0.01~pc. However, an increased mass interior to the Solar orbit will drag it inwards due to angular momentum conservation. The minimum angular momentum of the innermost stable circular orbit is $L_\mathrm{ISCO}=\sqrt{3}\, r_\mathrm{Schw} c$, whereas the Solar angular momentum is $\sim 2\times10^6\,\mathrm{pc\,km\,s}^{-1}$. These two quantities become equal when $M_\mathrm{BH}\simeq 4\times10^{13}\,M_\odot$\footnote{We are aware that this exceeds not only the mass of the Milky Way ($\sim 10^{12}\,M_\odot$), but of the entire Local Group (a few times larger; see \citealt{Strigari2025}). However, we are sure that intrepid theorists will find a mechanism that could support unabated growth of the SMBH mass as per our scenario.}, which corresponds to a time of less than $10^9$ years in the future. This is long compared to other impending disasters such as the AI takeover or the arrival of aliens disguised as interstellar comets, but shorter than anticipated extent of Earth habitability \citep{Ward2003} or the cosmological Big Rip (from which we were saved by a similarly linear evolution of the dark energy equation of state discovered by \citealt{DESI2025} on the April Fools day). Still, the problem lies sufficiently far in the future that the only immediate consequence would be the inability of future generation of astronomers to study the orbits of S-stars on a human timescale, unless the brave new world sees a dramatic increase in human lifespan.

\section{The rotation speed of the Galactic bar}  \label{sec:bar}

\begin{figure}
\includegraphics[width=\textwidth]{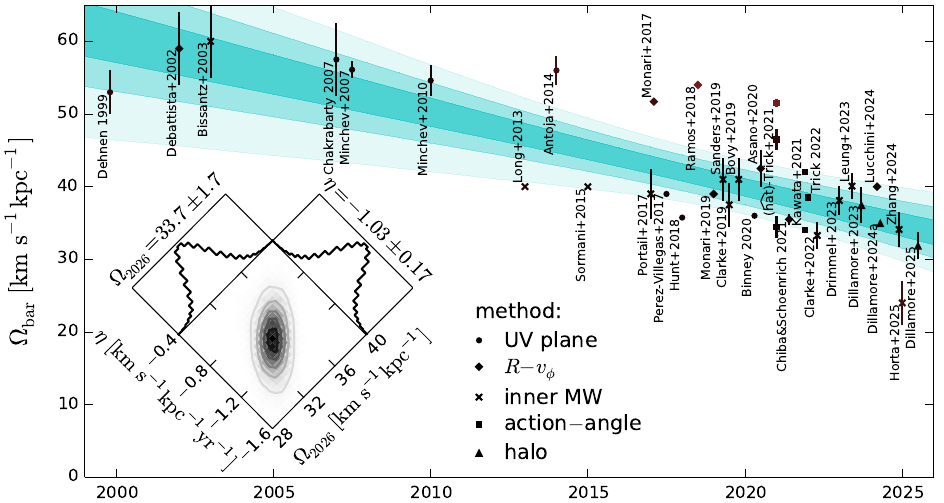}
\caption{Compilation of bar pattern speed measurements (an extended version of Figure~10 from \citealt{Hunt2025}). An unmistakable downward trend is quantified using the same approach as in Section~\ref{sec:mbh}, and a diagonal corner plot of the two most important parameters -- current pattern speed $\Omega_{2026}$ and slowdown rate $\eta$ -- is tucked into the left corner.
}  \label{fig:bar}
\end{figure}

We next turn our attention to the central region of the Milky Way outside the nuclear cluster, namely the Galactic bar, and focus on the estimates of its pattern speed (angular velocity) $\Omega_\mathrm{bar}$. Figure~\ref{fig:bar} shows a compilation of measurements since the beginning of this millenium, obtained by various methods. Again, some readers may congratulate themselves by noticing the same trend as we did, namely that $\Omega_\mathrm{bar}$ declines with time. A quantitative analysis of the same kind as in the previous section confirms this finding: the slowdown rate $\eta \equiv \mathrm{d}\Omega_\mathrm{bar}/\mathrm{d}t \approx -1\,\mathrm{km\,s}^{-1}\,\mathrm{kpc}^{-1}\,\mathrm{yr}^{-1}$.

That the bar slows down is not by itself surprising: it is well-known that bars experience dynamical friction against the DM\footref{footnote:dm2} halo \citep{Debattista2000,Athanassoula2003}. What is surprising is the very high deceleration rate; however, if we recall that a core-collapsed SIDM halo is a natural explanation of the SMBH mass growth, its extreme density may also lead to the rapid braking of the bar. In fact, a number of recent studies already presented an observational evidence for deceleration \citep{Chiba2021a,Zhang2025}; however, their estimates of $\eta$ are $\sim 10^9$ times lower, which may be attributed to a simple mix-up of units (a common source of errors even among experienced researchers).

\begin{figure}
\includegraphics[width=8cm]{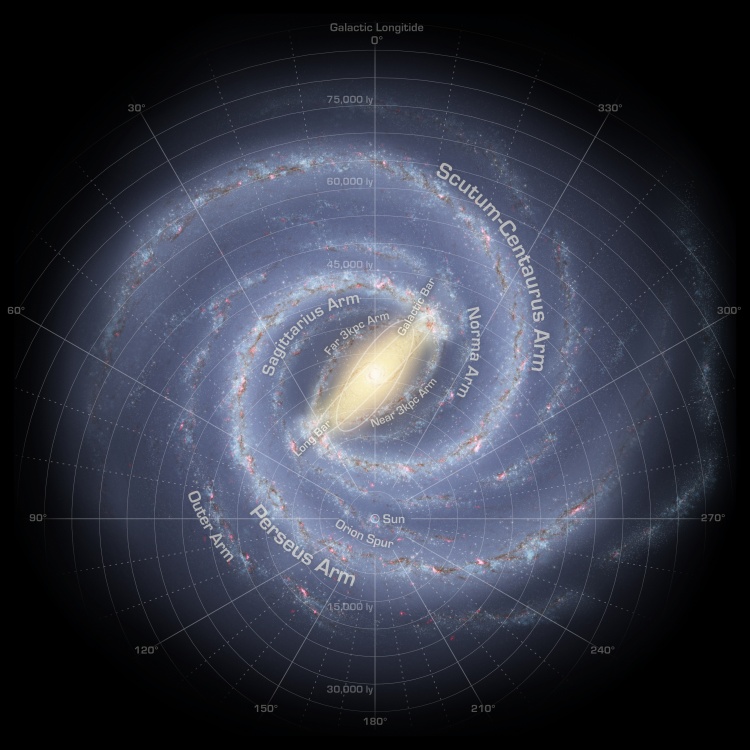}\includegraphics[width=8cm]{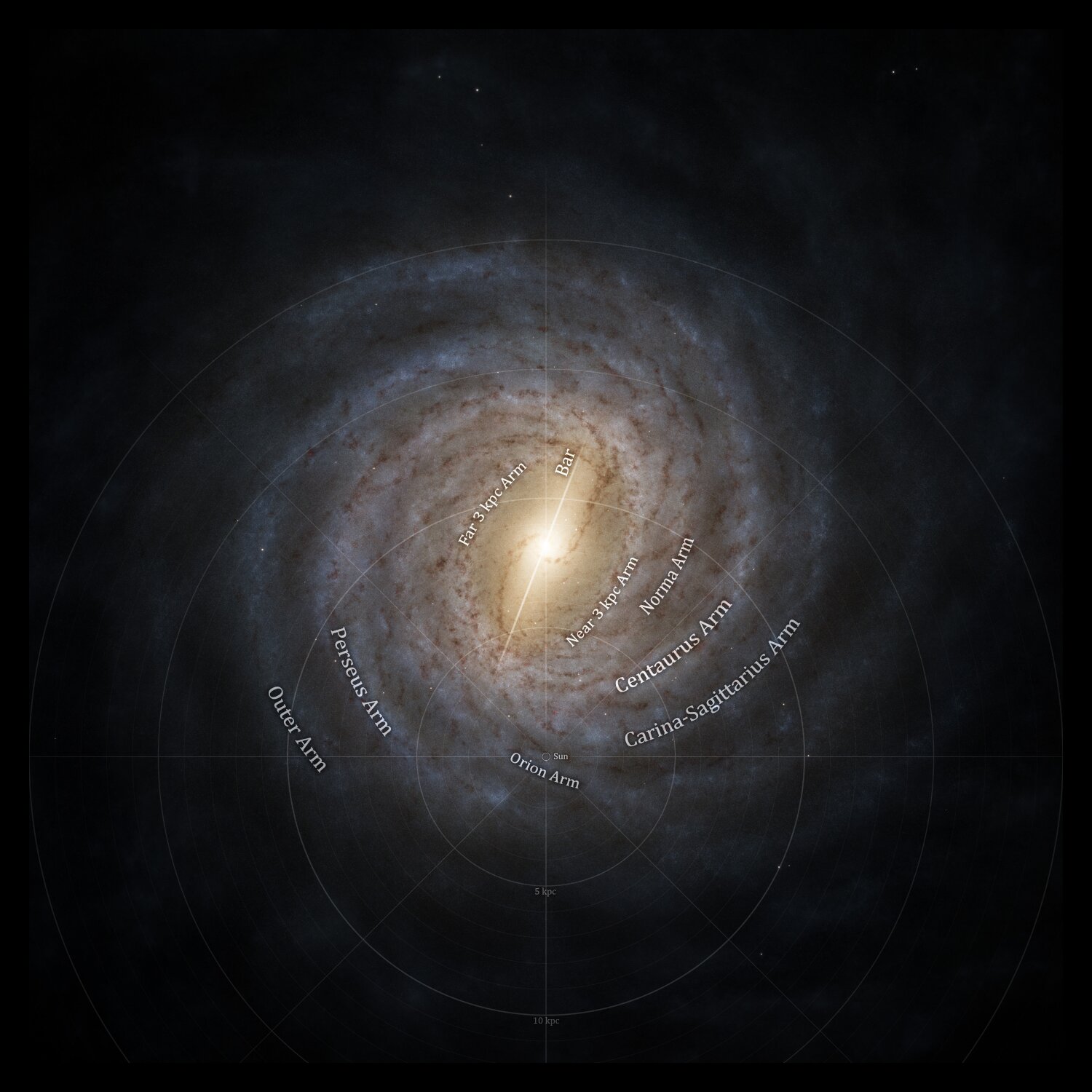}%
\begin{tikzpicture}[overlay]
\node[white] at(-15.2, 7.5) {\Large 2008};
\node[white] at( -0.8, 7.5) {\Large 2023};
\node[gray]  at(-14.5, 0.2) {\tiny\href{https://apod.nasa.gov/apod/ap080606.html}{[source: NASA / R.Hurt]}};
\node[gray]  at(- 2.1, 0.2) {\tiny\href{https://www.esa.int/ESA_Multimedia/Images/2023/12/Top-down_view_of_the_Milky_Way_annotated}{[source: ESA / S.Payne-Wardenaar]}};
\fill[red] (-12.0, 2.47) circle (0.04);
\draw[->,green] (-12.0, 2.47) .. controls (-12.4,2.47) and (-13.0,2.6) .. (-13.3, 3.1);
\node[green,rotate=-35,scale=0.75] at (-12.95,2.90) {\tiny Galaxy};
\node[green,rotate=-15,scale=0.75] at (-12.33,2.63) {\tiny rotation};
\fill[red] (-4.0, 2.46) circle (0.04);
\end{tikzpicture}%
\caption{Evolution of the Milky Way bar. The two images, taken 15 years apart, are on the same scale; the apparent mismatch in size is due to difference in exposure time. More important is the change in the bar angle, from $\sim 35^\circ$ on the left to $20^\circ$ on the right, amounting to the CCW rotation rate of $1^\circ\mathrm{yr}^{-1}$ in the opposite sense to the general CW Galactic rotation.
}  \label{fig:bar_angle}
\end{figure}

In addition to the pattern speed, many studies also attempted to measure the position angle of the bar (the angle between its major axis and the line connecting the Sun with the Galactic centre). As the bar (still) rotates faster than the angular velocity of the Sun ($\sim 30\,\mathrm{km\,s}^{-1}\,\mathrm{kpc}^{-1}$), we could expect that more recent observations would see it moving further in the direction of Galactic rotation. However, Figure~\ref{fig:bar_angle} reveals the opposite\footnote{These images are artistic impressions, but are based on real data (e.g., Figure~16 in \citealt{Churchwell2009}, Figure~17 in \citealt{Wegg2013}, Figure~15 in \citealt{Queiroz2021}, Figure~6 in \citealt{Zhang2024}, etc.)}: the angle has decreased from $35^\circ$ to $20^\circ$ in just 15 years! Not only this is $\sim10^7$ times faster than could be expected from the difference in angular velocities\footnote{This could be accounted for by introducing a suitable fudge factor, e.g., as in the reverberation mapping method.}; the sign is also wrong. Taken face value, this would indicate that the bar is travelling back in time!

We conjecture that the very fast inflow of DM during the core collapse, needed to explain the rapid growth of the SMBH mass and the equally rapid bar slowdown, may also create some relativistic effects, distorting the time measurement and even flipping its sign.
In fact, this is not the first time in recent history that unusual $\stackrel{\dots}{\textrm{u}}$ber-relativistic effects have been postulated. In 2011, the OPERA collaboration \href{https://arxiv.org/abs/1109.4897v1}{announced} a measurement of neutrino speed that was a tiny bit faster than light. Unfortunately, they had to retract their statement after a few months in the published version of the paper \citep{Opera2012}, but chronologically this gap is consistent with the apparent backward motion of the bar.

\textit{-- ``Sorry, we do not serve tachyons here''}\textsl{, says the bartender.\\
\indent A neutrino walks into a bar.}\footnote{This is a real joke of the era, not made up by some LLM.}\\
Indeed, if \textit{the bar walks into the Milky Way} at just the right time, it may travel backward!\\
The epoch of bar formation has been variously dated back to 3--8 Gyr ago \citep{Bovy2019, Haywood2024, Nepal2024, Sanders2024}, but given the above mentioned astronomically large fudge factors needed to reconcile the measurements with reality\footnote{Coincidentally, they may also explain the Hubble tension \citep[e.g.,][]{Shah2021,CervantesCota2023}, though we have not pursued this question any further.}, this may not be inconsistent with the OPERA result. In fact, we are lucky to still observe the bar at all: the growing SMBH will imminently destroy it in the relatively near future (once its mass exceeds a few percent of the disc mass\footnote{See \cite{Shen2004, Athanassoula2005, Hozumi2005} for the mechanism of bar destruction by a central mass concentration, although \cite{Wheeler2023} present an alternative point of view.}, i.e., in $\sim10^4$~yr). However, given the measured slowdown rate $\eta$, by this time the bar pattern speed will reach negative $10^4\,\mathrm{km\,s}^{-1}\,\mathrm{kpc}^{-1}$, which may dramatically affect the evolution of the entire disc due to the propeller effect, or even prevent the continued accretion of DM onto the SMBH. Apparently, more work is needed to establish \href{https://xkcd.com/3084/}{who will win this competition}.

\section{The distance to the Galactic Centre and the Solar migration}  \label{sec:migration}

\begin{figure}
\includegraphics[width=\textwidth]{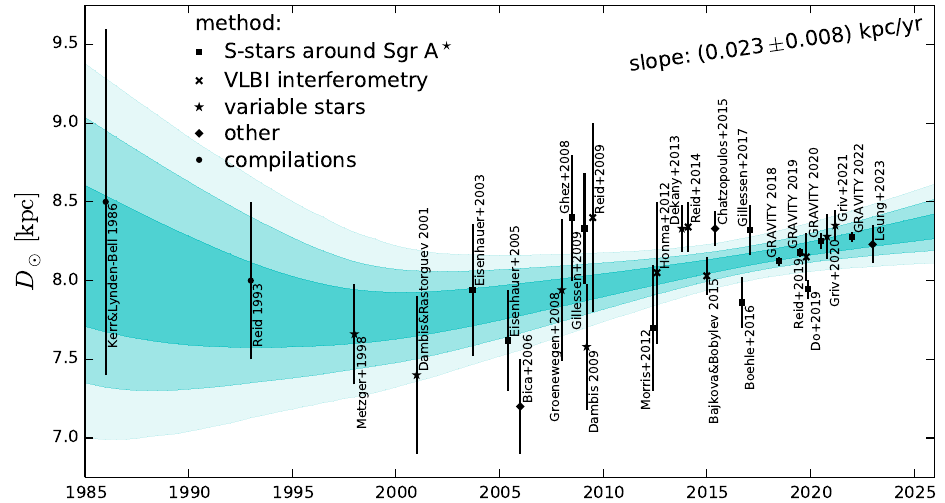}
\caption{Compilation of measurements of the distance from the Sun to the Galactic centre over the last four decades. A broken-linear trend line is fitted using the same likelihood approach as in Section~\ref{sec:mbh}, demonstrating that the earlier inward motion of the Sun has been replaced by an outward migration since late 1990s.
}  \label{fig:distance}
\end{figure}

Finally, we consider another important quantity that has less to do with the Milky Way as a whole, but more with our special position within it: namely, the distance from the Sun to the Galactic centre $D_\odot$. We would naively expect it to be slowly oscillating between peri- and apocentre radii ($\sim8$ and 10~kpc respectively) over the radial orbital period ($\sim180$~Myr)\footnote{These values are computed using the \texttt{MilkyWayPotential2022} model from tne \texttt{gala} dynamics package \citep{PriceWhelan2017}, which fits a variety of recent observational constraints (see section~7.7 in \citealt{Hunt2025}).}, but observations tell a different story.

Figure~\ref{fig:distance} shows a compilation of measurements of $D_\odot$ over the last 40 years. We note that a value $D_\odot=10$~kpc was adopted by IAU in 1964, then revised down to 8.5~kpc in 1985. This downward trend continues at a rate of $\sim$60~pc per year until late 1990s, and then appears to be reversed\footnote{While it is tempting to interpret this as evidence of the recent pericentre passage of the Sun, the time interval since the apocentre (less than 40 yr) is somewhat shorter than the radial orbital period computed by \texttt{gala}, so we are inclined to discard this hypothesis in favour of a more plausible one described below.}, so it is appropriate to fit a piecewise-linear model to these data. 
The inward motion is qualitatively consistent with the radial component of the Solar velocity, measured to be $v_R\simeq -10$\,km\,s$^{-1}$ \citep{Dehnen1998}, although large uncertainties prevent a more quantitative comparison. By contrast, the current rate of outward motion is constrained to be $\sim$23~pc\,yr$^{-1}$ at a 3$\sigma$ level of significance. A reader fluent in astronomical unit conversions would immediately notice that this is 70 times faster than the speed of light, which might raise some eyebrows. However, we stress that, by construction, $v_R$ is measured w.r.t.\ the local standard of rest, so if the entire Solar neighbourhood moves away from the Galactic centre with the same speed, this would be imperceptible to the observer relying on local kinematics. The apparent contradiction with special relativity should not be a concern, as explained in the previous section.

What could have caused these large-scale migrations of stars in the Solar neighbourhood? The inward motion can be attributed to the aforementioned core collapse of the SIDM halo and the associated increase of mass interior to the Solar orbit, coupled with the conservation of angular momentum. On the other hand, the transition to outward motion is likely caused by the dramatic slowdown of the bar and the corresponding migration of stars trapped in a resonance region \citep{Chiba2021a, Dillamore2024b, Zhang2025}.
Making very crude assumptions about the circular $\Omega_0$ and epicycle $\kappa$ frequencies in the disk, which may be extremely poorly known given the above arguments, we can briefly assume that the outer Lindblad resonance ($\Omega_P = \Omega_0 + \kappa /2$) went in a few years from about 6 kpc to 12 kpc. During that time frame, it could well be that the Sun was dramatically affected for a period of time, if not totally swept out. At this pace, the corotation resonance will soon catch up and move the Solar system out to even larger distances from the Galactic centre, thus preventing us from being consumed by the SMBH, to everyone's satisfaction.

\section{Discussion and conclusions}

Since ancient times, humans were fascinated with the orderly motion of celestial bodies, occasionally disturbed by a stray comet or a \href{https://en.wikipedia.org/wiki/Guest_star_(astronomy)}{guest star}. The millenia-long history of the Solar system observations culminated in the discovery of Kepler's laws, and ultimately led to the foundation of modern physics by Newton and colleagues. Although the heavens were considered immutable for much of human history, we now know that the Universe is not static (\citealt{Hubble1929, Lemaitre1931}, but see \citealt{Hoyle1948} for an alternative view and \citealt{Trimble2013} for a historical perspective). Its individual building blocks, namely galaxies, assemble and evolve over cosmic times, but still appear to be close to equilibrium at any given moment\footnote{The classic textbook \textit{Galactic Dynamics} by \citet{Binney2008} is largely devoted to equilibrium models, and in the context of this research note, might better be called \textit{Galactic Statics} (C.Hamilton, priv.comm.)}. However, as we learn more about the Milky Way thanks to the Gaia mission \citep{Gaia2016} and numerous other surveys, it became increasingly clear that our Galaxy is alive and kicking, with much of community's attention in recent years devoted to various disequilibrium effects in the Milky Way (see \citealt{Hunt2025,Frankel2026,Perryman2026} for reviews). Nevertheless, until now the dominant viewpoint has been that this evolution takes place on timescales of $10^6-10^9$ years. In our work, we show that this perception needs to be reconsidered.

In the preceding sections, we have used rigorous data analysis practices that are becoming increasingly more widespread in the astronomical community \citep[e.g.][]{Saha2003,BailerJones2017,Ting2025} to reveal unambiguous trends in the time evolution of key Galactic parameters over mere decades: the mass of the central SMBH, the pattern speed of the bar, and the distance from our Sun to the Galactic centre. At first glance these may appear to be unrelated\footnote{Of course, we cannot exclude more mundane explanations, such as some instrumental calibration effects (see table~3 in \citealt{Gravity2021} and section~6.2 in \citealt{Gillessen2009}), but these still would not account for the bar slowdown. Moreover, it is unlikely that such a mishap would close the window to new physics for the second time (after the OPERA result).}, but a deeper theoretical insight suggests a possible underlying physical mechanism that may be responsible for all three effects: the core collapse of the SIDM halo that just happens to unfold before our eyes. Whether this precise alignment in timing is coincidental or is a manifestation of some fundamental principle\footnote{We are not talking about intelligent design, mind you! Perhaps cosmological fine-tuning in the multiverse?} remains to be seen.

We hope that our analysis will draw attention of our colleagues to further explore these phenomena, but given their potentially far-reaching implications and to avoid obtrusive media attention, we opted to remain semi-anonymous. Nevertheless, if our findings prove to be correct, the N-committee would surely find a way to commemorate the impact of this groundbreaking work.


\end{document}